\documentstyle[psfig,pra,aps]{revtex}
\begin{document}
\draft
\title{Matter-field theory of the Casimir force 
\thanks{To be published in Phys. Rev. A
}}
\author{Masato Koashi$^1$ and Masahito Ueda$^2$}
\address{$^1$NTT Basic Research Laboratories, 3-1 Morinosato Wakamiya, 
Atsugi, Kanagawa 243-0198, Japan\\
$^2$Department of Physical Electronics, Hiroshima University,
Higashi-Hiroshima 739-8527, Japan}
\maketitle
\begin{abstract}
A matter-field theory of the Casimir force is formulated in which the 
electromagnetic field and collective modes of dielectric media
are treated on an equal footing. In our theory, the Casimir force is 
attributed to zero-point energies of the combined matter-field modes.
We analyze why some of the existing theories favor the interpretation
of the Casimir force as originating from zero-point energies of the
electromagnetic field and others from those of the matter.
\end{abstract}


\section{INTRODUCTION}
It is well known that two conducting neutral plates placed at a small 
distance attract each other. This phenomenon is 
often called the Casimir effect because of the
celebrated formula he derived for the force between two perfectly
conducting plates\cite{casimir48}.
Casimir calculated the sum of the quantum-mechanical zero-point energies
of the normal modes of the electromagnetic (EM) field between 
the two metal plates,
and showed that the total energy depends on the distance between the 
plates. The spatial derivative of the sum
gives what we now call the Casimir force.
 The generalization of the Casimir effect 
to the case of two dielectric plates
was made by Lifshitz\cite{lifshitz56}, and his formula 
was rederived by van Kampen, Nijboer, and Schram\cite{kampen68,schram73} 
using a technique attributed to
Casimir. The Casimir force is essentially 
a long-range van der Waals force, 
where one cannot ignore the delay caused by the finiteness of the 
velocity of light. This retardation effect is not always a small 
correction and may alter the very nature of the force. For 
example, the Casimir force between two perfectly conducting 
hemispherical shells has been shown to be repulsive\cite{boyer68}. 

Experimental studies of the Casimir force started 
long ago\cite{sparnaay58,tabor68,israelachvili72}
and have culminated in a recent high-resolution measurement
by Lamoreaux\cite{lamoreaux97}.
A closely related phenomenon, the retarded attraction force between 
an atom and a metal plate (the Casimir-Polder force\cite{polder48}),
was also measured successfully by Sukenik {\it et al}.\cite{sukenik93}.
With further improvements, it will hopefully be possible to confirm
experimentally the theoretical predictions of
several corrections, such as finite-temperature effects
~\cite{mehra67,brown69,schwinger78} 
and  radiative corrections\cite{xinwei97}.

The scope of the research of the Casimir effect covers
 many areas of physics and other
fields, ranging  from biology to cosmology and elementary particle physics.
Reviews and extensive references are available, for example, in 
Refs.~\cite{plunien85,mostepanenko88,elizalde91,spruch96}.

The Casimir effect has usually been attributed to 
zero-point fluctuations of the EM field.
However, Schwinger and his collaborators\cite{schwinger78,schwinger75} 
showed that the Casimir effect can be 
derived in terms of 
Schwinger's source theory, which has no explicit reference to
vacuum-field fluctuations of the 
EM field. 
Milonni and Shih have recently developed a source theory of the 
Casimir effect within 
the framework of 
conventional quantum electrodynamics \cite{milonni92}. 
In this theory, the Casimir force originates from
quantum fluctuations of atomic dipoles in the dielectric, and 
the EM field plays only the passive role of mediating
interactions between those dipoles.

It seems like it is only a matter of taste whether we attribute
the Casimir effect to the quantum nature of the EM field 
or to that of the matter. Accordingly, 
the following questions naturally arise: Do the field and the matter 
really play (or to what extent do they play) symmetrical roles in this problem? 
Why do some approaches emphasize the quantum nature of the EM field 
and others stress that of the matter?
To answer these questions, it would be of interest to discuss this 
problem from a standpoint that has no preference for the field or for the 
matter. Unfortunately, existing theories do not suit this purpose because 
they all invoke in one way or another the Maxwell equations in which 
the degrees of freedom of the matter are in advance embedded in the
frequency dependence of the dielectric response function.

The primary purpose of this paper is to propose a matter-field theory
of the Casimir force in which the matter and 
the field are treated on an equal footing. 
Our strategy is to explicitly diagonalize the matter-field Hamiltonian
which is quadratic in its dynamical variables. Because relevant physical
quantities are then expressed in terms of eigenvalues and eigenvectors of
the full Hamiltonian, all physical effects allow unambiguous interpretation.

This paper is organized as follows.
Section~\ref{sec:formal} shows how the Casimir force is derived, 
starting from the Lagrangian describing 
the interaction between the EM field and collective 
modes of the matter.
Section~\ref{sec:max} proves that this 
derivation leads to the same results as existing 
theories~\cite{casimir48,lifshitz56,kampen68,schram73,schwinger78,schwinger75,milonni92}. 
It will be shown that in our treatment 
both field and matter contributes zero-point energies. 
Section~\ref{sec:previous} analyzes the previous derivations
of the Casimir force
and points out that an inherent asymmetry indeed exists 
in the problem between 
the field and the matter, which favors the
interpretation of the Casimir force as originating from
 zero-point fluctuations of
the EM field in some formulations\cite{kampen68,schram73},
and from those of the matter in others\cite{milonni92}.
Section \ref{sec:conclusion}  summarizes the main conclusions 
of this paper and describes our 
answers to the above questions.

\section{MATTER-FIELD THEORY}
\label{sec:formal}
\subsection{Formulation of the problem}
\begin{figure}
\begin{center}
\mbox{}
\psfig{file=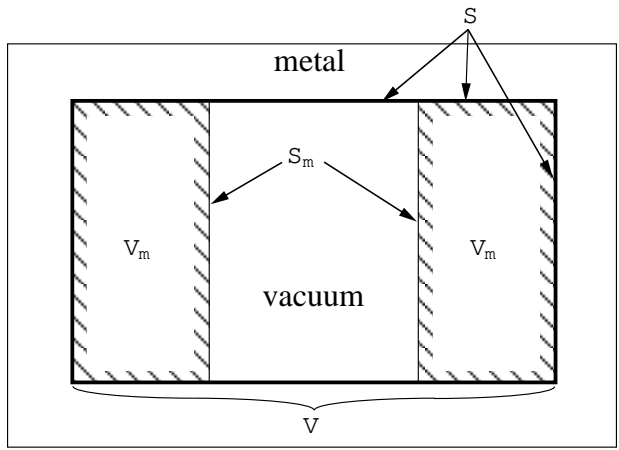,height=6cm}
\end{center}
\caption{Two dielectric slabs (region $V_m$) are separated by
the vacuum. The entire system $V$ is enclosed by a perfectly conducting metal.
The boundary of the system on the metal is denoted by $S$, and the
boundary between the dielectric slabs and the vacuum of the
EM field is denoted by $S_m$. 
\label{fig:1}
}
\end{figure}

Consider a system in which two dielectric slabs are separated
by the vacuum of the EM field as 
shown in Fig.\ \ref{fig:1}. We assume that the entire system $V$ 
is enclosed by a perfectly conducting metal, and denote 
the boundary of the system on the metal as $S$. 
The region $V_m$ occupied by the dielectric is 
surrounded by $S$ and by 
the interface $S_m$ between the dielectric and the vacuum. 
The following discussion does not depend on concrete
shapes of $V$ and $V_m$, but we require that any closed 
loop or any closed surface in $V$ can continuously shrink
to a point without going outside of $V$.
 The condition concerning closed loops ensures that 
any irrotational vector function
${\bf Q}({\bf r})$ [i.e., $\bbox{\nabla}\times{\bf Q}({\bf r})={\bf 0}$]
has a scalar potential $\phi({\bf r})$ with 
$-\bbox{\nabla}\phi({\bf r})={\bf Q}({\bf r})$.
The condition concerning closed surfaces ensures that 
if ${\bf Q}({\bf r})$ has a scalar potential and ${\bf Q}\perp S$,
${\bf Q}({\bf r})$ has a scalar potential $\phi$ that vanishes
everywhere on $S$. We impose the same requirements on 
the topology of $V_m$.

In a field theory of the Casimir force\cite{schram73}, 
normal-mode frequencies of 
the ``EM field'' are determined from the Maxwell equations with proper 
boundary conditions, and the degrees of freedom of the dielectric 
slabs do not appear explicitly because the effect of the presence 
of the dielectric is incorporated into the theory through the dielectric 
response function $\epsilon(\omega)$. 
In a matter theory of the Casimir force\cite{milonni92}, 
the Green function of the EM field,
which mediates the interaction between atomic dipoles,
 is determined from 
the same dielectric response function. 
Because our goal is to treat the EM field and the matter on an equal 
footing, we first consider the EM field and 
the dielectric media separately, and then introduce the interaction 
between them.

The Lagrangian $L_{\text{field}}$ of the free EM field is written as 
\begin{equation}
L_{\text{field}}=
\int_Vdv\left(\frac{\epsilon_0}{2}|{\bf E}|^2
-\frac{1}{2\mu_0}|{\bf B}|^2\right),
\end{equation}
where we assume that the EM field is only subject to a perfectly
conducting boundary $S$.

We take the following simple model for the dielectric: 
It consists of various kinds of particles whose species are labeled by $j$ 
with number density $n_j$. The $j$th species has mass 
$m_j$ and charge $e_j$, and particles belonging to this species
are bounded around their equilibrium positions by a restoring force 
characterized by frequency $\omega_j$. In order to describe their collective
 motions, we use 
a complete orthonormal set $\{ \bbox{\beta}_i({\bf r}) \}$ of 
mode functions
in the region $V_m$ that is 
occupied by the dielectric. They satisfy 
\begin{equation}
\int_{V_m}\bbox{\beta}_i({\bf r})\cdot\bbox{\beta}_j({\bf
 r})dv=\delta_{ij},
\end{equation}
where $\delta_{ij}$ is the Kronecker's delta.
We take the boundary condition 
on $S$ as $\bbox{\beta}_i\parallel S$ and that on $S_m$  
as 
$\bbox{\beta}_i\parallel S_m$. Note that imposing a boundary condition
does not alter the physics in our model because we assume no interaction
(except that mediated by the EM field) between the particles. 
Under this boundary condition, the set $\{ \bbox{\beta}_i \}$ is uniquely 
classified into the set of transverse vector fields $\{ \bbox{\beta}^{(T)}_i
 \}$
satisfying $\bbox{\nabla}\cdot\bbox{\beta}^{(T)}_i=0$, and the set of longitudinal 
ones $\{ \bbox{\beta}^{(L)}_i \}$ satisfying
 $\bbox{\nabla}\times\bbox{\beta}^{(L)}_i={\bf 0}$.

The polarization density ${\bf P}_j({\bf r})$ of the $j$th 
species can be expanded in terms of these bases as
\begin{equation}
{\bf P}_j({\bf r})=
\sqrt{\epsilon_0}\sum_i\bbox{\beta}^{(T)}_i({\bf r})X_{ij}^{(T)}
+\sqrt{\epsilon_0}\sum_i\bbox{\beta}^{(L)}_i({\bf r})X_{ij}^{(L)},
\end{equation}
where $X_{ij}^{(T)}$ and $X_{ij}^{(L)}$ are the generalized coordinates 
that describe the collective displacements of the $j$th species.
The corresponding current density ${\bf j}_j({\bf r})$ and the charge density 
$\rho_j({\bf r})$ are given by
\begin{equation}
{\bf j}_j({\bf r})=\dot{{\bf P}}_j({\bf r})=
\sqrt{\epsilon_0}\sum_i\bbox{\beta}^{(T)}_i({\bf r})\dot{X}_{ij}^{(T)}
+\sqrt{\epsilon_0}\sum_i\bbox{\beta}^{(L)}_i({\bf r})\dot{X}_{ij}^{(L)}
\label{eq:j}
\end{equation}
and
\begin{equation}
\label{eq:charge}
\rho_j({\bf r})=-\bbox{\nabla}\cdot{\bf P}_j({\bf r})=
-\sqrt{\epsilon_0}\sum_i\bbox{\nabla}\cdot\bbox{\beta}^{(L)}_i({\bf
 r})X_{ij}^{(L)}.
\end{equation}
In Eq.\ (\ref{eq:j}), we have neglected nonlinear terms in 
$X_{ij}^{(T,L)}$, which give rise to the term 
${\bf v}\times {\bf B}$ of the Lorentz force.
They may provide a relativistic correction to the Casimir force, which,
however, does not arise in the lowest nontrivial order we are 
concerned with.
The total kinetic energy of the $j$th particle is written as follows:
\begin{equation}
T_j  = 
\int_{V_m}dv\frac{1}{2}n_jm_j
\left(\frac{{\bf j}_j({\bf r})}{e_jn_j}\right)^2
=\sum_j\frac{1}{2\omega_{pj}^2}
(\sum_i\dot{X}_{ij}^{(T)2}+\sum_i\dot{X}^{(L)2}_{ij}),
\end{equation}
where $\omega_{pj}\equiv \sqrt{e_j^2 n_j/\epsilon_0 m_j}$ is the plasma
 frequency.
The Lagrangian for the dielectric $L_{\text{matter}}$ is thus given by
\begin{equation}
L_{\text{matter}}=
\sum_{ij}\frac{1}{2\omega_{pj}^2}(\dot{X}^{(T)2}_{ij}-\omega_{0j}^2X^{(T)2}_{ij})
+\sum_{ij}\frac{1}{2\omega_{pj}^2}(\dot{X}^{(L)2}_{ij}-\omega_{0j}^2X^{(L)2}_{ij}).
\end{equation}

The total Lagrangian $L$ for the whole system is the sum of 
$L_{\text{field}}$,
$L_{\text{matter}}$, and the interaction part:
\begin{equation}
L=L_{\text{field}}+L_{\text{matter}}
-\sum_j\int_{V_m}dv\left[\rho_j({\bf r})\varphi({\bf r})
-{\bf j}_j({\bf r})\cdot{\bf A}({\bf r})\right],
\end{equation}
where $\varphi({\bf r})$ and ${\bf A}({\bf r})$ are respectively 
the scalar and 
the vector potential of the EM field. The corresponding Hamiltonian
in the Coulomb gauge is derived as follows:
\begin{eqnarray}
H&=&\sum_{ij}\left(
\dot{X}^{(T)}_{ij}\frac{\partial L}{\partial\dot{X}^{(T)}_{ij}}
+\dot{X}^{(L)}_{ij}\frac{\partial L}{\partial\dot{X}^{(L)}_{ij}}
\right)
+\dot{\bf A}({\bf r})\cdot\frac{\delta L}{\delta \dot{\bf A}({\bf r}) }
-L
\nonumber \\
&=&\sum_{ij}\left[\frac{\omega_{pj}^2}{2}
(P^{(T)}_{ij}-\sqrt{\epsilon_0}\int_{V_m}\bbox{\beta}^{(T)}_i\cdot{\bf
 A}dv)^2
+\frac{\omega_{0j}^2}{2\omega_{pj}^2}X^{(T)2}_{ij}\right]
\nonumber \\
& & +\sum_{ij}\left[\frac{\omega_{pj}^2}{2}
(P_{ij}^{(L)}-\sqrt{\epsilon_0}\int_{V_m}\bbox{\beta}^{(L)}_i\cdot{\bf
 A}dv)^2
+\frac{\omega_{0j}^2}{2\omega_{pj}^2}X_{ij}^{(L)2}\right]
\nonumber \\
& & +\int_Vdv\left(\frac{\epsilon_0}{2}|{\bf E}^{(T)}|^2
+\frac{1}{2\mu_0}|{\bf B}|^2\right)
+\int_V\frac{\epsilon_0}{2}|{\bf E}^{(L)}|^2dv,
\label{eq:H1}
\end{eqnarray}
where ${\bf E}^{(T)}$ and ${\bf E}^{(L)}$ are respectively the transverse
 and 
the longitudinal parts of the electric field; 
$P_{ij}^{(T)}\equiv \partial L / \partial\dot{X}^{(T)}_{ij}$ and
 $P_{ij}^{(L)}\equiv \partial L / \partial\dot{X}^{(L)}_{ij}$ 
are the generalized momenta conjugate to $X_{ij}^{(T)}$ and $X_{ij}^{(L)}$,
respectively.

The transverse part of the EM field can be decomposed into 
the normal modes in the usual manner. Consider a complete orthonormal set 
 of 
transverse mode functions $\{ \bbox{\alpha}^{(T)}_i({\bf r}) \}$ in $V$
 that satisfy 
\begin{equation}
\triangle^2\bbox{\alpha}^{(T)}_i({\bf r})\
=-\frac{\Omega^2_i}{c^2}\bbox{\alpha}^{(T)}_i({\bf r}),
\label{eq:Omega}
\end{equation}
\begin{equation}
\int_{V}\bbox{\alpha}^{(T)}_i({\bf r})
\cdot\bbox{\alpha}^{(T)}_j({\bf r})dv=\delta_{ij}.
\end{equation}
Because $S$ is the boundary on the perfect conductor,
$\bbox{\alpha}^{(T)}_i$ should satisfy
the boundary condition $\bbox{\alpha}^{(T)}_i \perp S$. The transverse
fields ${\bf A}({\bf r})$ and ${\bf E}^{(T)}({\bf r})$ can be expanded with 
this base as
\begin{equation}
{\bf A}({\bf r})
=\frac{1}{\sqrt{\epsilon_0}}\sum_i\bbox{\alpha}^{(T)}({\bf r})q_i
\label{eq:A}
\end{equation}
and
\begin{equation}
{\bf E}^{(T)}({\bf r})
=-\frac{1}{\sqrt{\epsilon_0}}\sum_i\bbox{\alpha}^{(T)}({\bf r})p_i ,
\end{equation}
where $q_i$ and $p_i$ are canonically conjugate pairs.
The transverse EM field part of the Hamiltonian (\ref{eq:H1}) 
then reduces to a collection of harmonic oscillators: 
\begin{equation}
\int_Vdv\left(\frac{\epsilon_0}{2}|{\bf E}^{(T)}|^2
+\frac{1}{2\mu_0}|{\bf B}|^2\right)
=\sum_i\left(\frac{1}{2}p_i^2+\frac{1}{2}\Omega_i^2q_i^2\right).
\label{eq:fi-ene}
\end{equation}

In the Coulomb gauge,
the longitudinal part ${\bf E}^{(L)}({\bf r})$ of 
the electric field  
satisfies the Poisson equation 
$\epsilon_0\bbox{\nabla}\cdot{\bf E}^{(L)}=\rho$,
where the total charge $\rho$ is given from Eq.\ (\ref{eq:charge}) as
\begin{equation}
\rho=\sum_j\rho_j=
-\sqrt{\epsilon_0}\sum_{ij}\bbox{\nabla}\cdot\bbox{\beta}^{(L)}_i({\bf r})
X_{ij}^{(L)}
\end{equation}
in $V_m$ and zero elsewhere.
Let us define the longitudinal mode function $\bbox{\alpha}_i^{(L)}({\bf r})$
in $V$ that satisfies the boundary condition 
$\bbox{\alpha}_i^{(L)} \perp S$ and
\begin{equation}
\bbox{\nabla}\cdot\bbox{\alpha}_i^{(L)}({\bf r})=
\left\{ \begin{array}{ll}
\bbox{\nabla}\cdot\bbox{\beta}_i^{(L)}({\bf r}) \;\; & \mbox{in $V_m$}, \\
0  & \mbox{otherwise}.
\end{array} \right.
\label{eq:defalpha}
\end{equation}
The solution to Eq.\ (\ref{eq:defalpha}) is given as
\begin{equation}
\bbox{\alpha}_i^{(L)}({\bf r})=
\left\{ \begin{array}{lll}
\bbox{\beta}_i^{(L)}({\bf r}) &
 -\displaystyle{\sum_k\bbox{\alpha}_k^{(T)}({\bf r})
\int_{V_m}\bbox{\alpha}_k^{(T)}({\bf r}^\prime)\cdot\bbox{\beta}_i^{(L)}({\bf r}^\prime)dv^\prime
 } \;\; & \mbox{in $V_m$}, \\
& -\displaystyle{\sum_k\bbox{\alpha}_k^{(T)}({\bf r})
\int_{V_m}\bbox{\alpha}_k^{(T)}({\bf r}^\prime)\cdot\bbox{\beta}_i^{(L)}({\bf r}^\prime)dv^\prime 
} & \mbox{otherwise}.
\end{array} \right.
\label{eq:alphal}
\end{equation}
Note that $\bbox{\alpha}_i^{(L)}({\bf r})$ is neither normalized 
nor orthogonal. 
With this base and $X^{(L)}_{ij}$,
${\bf E}^{(L)}({\bf r})$ can now be expanded as follows:
\begin{equation}
{\bf E}^{(L)}({\bf r})=-\frac{1}{\sqrt\epsilon_0}
\sum_{ij}\bbox{\alpha}_i^{(L)}({\bf r})
X_{ij}^{(L)}.
\end{equation}
The Coulomb-energy part of the Hamiltonian (\ref{eq:H1}) is 
\begin{equation}
\int_V\frac{\epsilon_0}{2}|{\bf E}^{(L)}|^2dv=\frac{1}{2}\sum_{ik}
\left(\sum_jX_{ij}^{(L)}\right)
\left(\sum_{j^\prime} X_{kj^\prime}^{(L)}\right)
\int_V\bbox{\alpha}_i^{(L)}({\bf r})\cdot\bbox{\alpha}^{(L)}_k({\bf r})dv.
\label{eq:co-ene}
\end{equation}
Using Eq.\ (\ref{eq:alphal})
and the orthogonality between $\bbox{\alpha}_i^{(T)}({\bf r})$ 
and $\bbox{\alpha}^{(L)}_k({\bf r})$,
 the integral on the right-hand side 
of Eq.\ (\ref{eq:co-ene}) can be rewritten 
as follows:
\begin{equation}
\int_V\bbox{\alpha}_i^{(L)}({\bf r})\cdot\bbox{\alpha}^{(L)}_k({\bf r})dv
=\int_{V_m}\bbox{\beta}_i^{(L)}({\bf r})
\cdot\bbox{\alpha}^{(L)}_k({\bf r})dv.
\label{eq:int}
\end{equation}
Combining Eqs. (\ref{eq:H1}), (\ref{eq:A}), (\ref{eq:fi-ene}),
(\ref{eq:co-ene}), and (\ref{eq:int}), we obtain the following expression for
 the 
total Hamiltonian:
\begin{eqnarray}
H&=&\sum_{ij}\left[\frac{\omega_{pj}^2}{2}
\left(P^{(T)}_{ij}-\sum_kq_k
\int_{V_m}\bbox{\alpha}^{(T)}_k\cdot\bbox{\beta}^{(T)}_idv\right)^2
+\frac{\omega_{0j}^2}{2\omega_{pj}^2}X^{(T)2}_{ij}\right]
\nonumber \\
& & +\sum_{ij}\left[\frac{\omega_{pj}^2}{2}
\left(P^{(L)}_{ij}-\sum_kq_k
\int_{V_m}\bbox{\alpha}^{(T)}_k\cdot\bbox{\beta}^{(L)}_idv\right)^2
+\frac{\omega_{0j}^2}{2\omega_{pj}^2}X^{(L)2}_{ij}\right]
\nonumber \\
& & +\sum_i\left(\frac{1}{2}p_i^2+\frac{1}{2}\Omega_i^2q_i^2\right)
\nonumber \\
& & +\frac{1}{2}\sum_{ik}
\left(\sum_jX_{ij}^{(L)}\right)
\left(\sum_{j^\prime} X_{kj^\prime}^{(L)}\right)
\int_{V_m}\bbox{\beta}_i^{(L)}({\bf r})\cdot\bbox{\alpha}^{(L)}_k({\bf
 r})dv.
\label{eq:H2}
\end{eqnarray}

\subsection{Plasmons}
In this section, we discuss the physical implications of 
the longitudinal displacement represented by 
$\{\bbox{\beta}^{(L)}_i({\bf r})\}$. Let us consider another
set of complete orthonormal modes 
$\{\tilde{\bbox{\beta}}_i({\bf r})\}$ in $V_m$ that satisfy 
different boundary conditions on the boundaries, namely, 
$\tilde{\bbox{\beta}}_i \perp S$ and $\tilde{\bbox{\beta}}_i \perp S_m$. 
Under these new boundary conditions,
the set $\{ \tilde{\bbox{\beta}}_i \}$ is uniquely 
separated into the set of transverse vector fields 
$\{ \tilde{\bbox{\beta}}^{(T)}_i \}$
satisfying $\bbox{\nabla}\cdot\tilde{\bbox{\beta}}^{(T)}_i=0$, 
and the set of longitudinal 
ones $\{ \tilde{\bbox{\beta}}^{(L)}_i \}$, which have 
scalar potentials $\tilde{\phi}_i({\bf r})$
($\tilde{\bbox{\beta}}^{(L)}_i=-\bbox{\nabla}\tilde{\phi}_i$)
 that vanish
on $S$ and $S_m$.

The orthogonality between 
$\tilde{\bbox{\beta}}_k^{(L)}$ and $\bbox{\beta}_i^{(T)}$ is shown as follows:
\begin{equation}
\int_{V_m}\bbox{\beta}_i^{(T)}({\bf r})
\cdot\tilde{\bbox{\beta}}_k^{(L)}({\bf r})dv
=-\int{\bf n}\cdot\bbox{\beta}_i^{(T)}({\bf r})
\tilde{\phi}_k({\bf r})dS
=0,
\end{equation}
where the surface integral runs over the surface of $V_m$, and 
${\bf n}$ is the unit vector normal to that surface. In contrast,
$\tilde{\bbox{\beta}}_i^{(T)}$ and $\bbox{\beta}_k^{(L)}$ are not
necessarily orthogonal. This implies that a vector longitudinal 
 under the boundary conditions $\bbox{\beta}_i\parallel S,S_m$
 is not necessarily longitudinal
under the boundary conditions $\tilde{\bbox{\beta}}_i\perp S,S_m$.
Let us choose the base 
$\{\bbox{\beta}_i^{(L)}\}$ as a union of two such orthogonal sets,
$\{\bbox{\beta}_i^{(b)}\}$ and $\{\bbox{\beta}_i^{(s)}\}$, that 
$\bbox{\beta}_i^{(b)}$ can be expanded with 
$\{\tilde{\bbox{\beta}}_i^{(L)}\}$, and $\bbox{\beta}_i^{(s)}$ with 
$\{\tilde{\bbox{\beta}}_i^{(T)}\}$. 
Associated with $\bbox{\beta}_i^{(b)}$ and $\bbox{\beta}_i^{(s)}$,
we may introduce 
$\bbox{\alpha}_i^{(b)}$ and $\bbox{\alpha}_i^{(s)}$
through the relation (\ref{eq:alphal}).

Since $\bbox{\beta}_i^{(b)}$ 
has a scalar potential $\phi^{(b)}_i$ 
($\bbox{\beta}^{(b)}_i=-\bbox{\nabla}\phi^{(b)}_i$)
that vanishes
on the boundaries
$S$ and $S_m$, part of the integral (\ref{eq:int}) that includes 
the $\bbox{\beta}_i^{(b)}$ mode can be further simplified to be
\begin{eqnarray}
\int_{V_m}\bbox{\beta}_i^{(b)}({\bf r})\cdot\bbox{\alpha}^{(L)}_k({\bf
 r})dv
&=&\int_{V_m}\phi_i^{(b)}({\bf r})\bbox{\nabla}\cdot\bbox{\alpha}^{(L)}_k({\bf
 r})dv
=\int_{V_m}\phi_i^{(b)}({\bf r})\bbox{\nabla}\cdot\bbox{\beta}^{(L)}_k({\bf r})dv
\nonumber \\
&=&\int_{V_m}\bbox{\beta}_i^{(b)}({\bf r})\cdot\bbox{\beta}^{(L)}_k({\bf
 r})dv,
\end{eqnarray}
where we used Eq.\ (\ref{eq:defalpha}) in deriving the second equality.
Because of the orthogonality condition between 
$\bbox{\beta}_i^{(b)}$ and $\bbox{\beta}_k^{(s)}$, we obtain
\begin{equation}
\int_{V_m}\bbox{\beta}_i^{(b)}({\bf r})\cdot\bbox{\alpha}^{(s)}_k({\bf
 r})dv
=0
\end{equation}
and because of the orthonormality of $\{\bbox{\beta}_i^{(b)}\}$,
we obtain
\begin{equation}
\int_{V_m}\bbox{\beta}_i^{(b)}({\bf r})\cdot\bbox{\alpha}^{(b)}_k({\bf
 r})dv
=\delta_{ik}.
\end{equation}
From these, the static 
Coulomb-interaction part of the Hamiltonian (\ref{eq:H2})
can be decomposed into two parts:
\begin{eqnarray}
&&\frac{1}{2}\sum_{ik}
\left(\sum_jX_{ij}^{(L)}\right)
\left(\sum_{j^\prime} X_{kj^\prime}^{(L)}\right)
\int_{V_m}\bbox{\beta}_i^{(L)}({\bf r})\cdot\bbox{\alpha}^{(L)}_k({\bf
 r})dv
\nonumber \\
&=&\frac{1}{2}\sum_{ik}
\left(\sum_jX_{ij}^{(s)}\right)
\left(\sum_{j^\prime} X_{kj^\prime}^{(s)}\right)
\int_{V_m}\bbox{\beta}_i^{(s)}({\bf r})\cdot\bbox{\alpha}^{(s)}_k({\bf
 r})dv
+\frac{1}{2}\sum_{i}
\left(\sum_jX_{ij}^{(b)}\right)^2.
\label{eq:coulomb}
\end{eqnarray}

This result shows that the Coulomb interaction between the
modes $X^{(b)}_{ij}$ does not depend on the boundaries (bulk plasmons).
Further, these modes do not interact with the transverse EM
field since
\begin{equation}
\int_{V_m}\bbox{\alpha}_i^{(T)}({\bf r})\cdot\bbox{\beta}^{(b)}_k({\bf
 r})dv
=-\int{\bf n}\cdot\bbox{\alpha}_i^{(T)}({\bf r})\phi^{(b)}_k({\bf r})dS
=0,
\label{eq:vanish}
\end{equation}
where the surface integral runs over the surface of $V_m$.
The bulk plasmons thus do not contribute to the Casimir force.

The modes $X^{(s)}_{ij}$ describe the surface plasmons 
whose resonance frequencies do depend on the parameters of the surface 
of $V_m$ and
its outside, through the integral in Eq.\ (\ref{eq:coulomb}). The surface
 plasmons
are also affected by the surroundings through the interaction with the 
transverse EM field, for their scalar potentials by definition should not 
vanish everywhere on the surface of $V_m$ and 
the integral $\int_{V_m}\bbox{\alpha}_i^{(T)}\bbox{\beta}^{(s)}_kdv$
does not vanish unlike 
Eq.\ (\ref{eq:vanish}).

\subsection{Combined oscillations}

After separating the longitudinal modes into the two types of plasmons,
the total Hamiltonian is now written as follows:
\begin{eqnarray}
H&=&\sum_{ij}\left[\frac{\omega_{pj}^2}{2}
\left(P^{(T)}_{ij}-\sum_kq_k
\int_{V_m}\bbox{\alpha}^{(T)}_k\cdot\bbox{\beta}^{(T)}_idv\right)^2
+\frac{\omega_{0j}^2}{2\omega_{pj}^2}X^{(T)2}_{ij}\right]
\nonumber \\
& & +\sum_{ij}\left[\frac{\omega_{pj}^2}{2}
\left(P^{(s)}_{ij}-\sum_kq_k
\int_{V_m}\bbox{\alpha}^{(T)}_k\cdot\bbox{\beta}^{(s)}_idv\right)^2
+\frac{\omega_{0j}^2}{2\omega_{pj}^2}X^{(s)2}_{ij}\right]
\nonumber \\
& & +\frac{1}{2}\sum_{ik}K_{ik}
\left(\sum_jX_{ij}^{(s)}\right) 
\left(\sum_{j^\prime} X_{kj^\prime}^{(s)}\right) 
\nonumber \\
& & +\sum_i\left(\frac{1}{2}p_i^2+\frac{1}{2}\Omega_i^2q_i^2\right)
\nonumber \\
& & +\sum_{ij}\left(\frac{\omega_{pj}^2}{2}
P^{(b)2}_{ij}
+\frac{\omega_{0j}^2}{2\omega_{pj}^2}X^{(b)2}_{ij}\right)
+\frac{1}{2}\sum_{i}
\left(\sum_jX_{ij}^{(b)}\right)^2,
\label{eq:H3}
\end{eqnarray}
where
\begin{equation}
K_{ik}\equiv
\int_{V_m}\bbox{\alpha}_i^{(s)}({\bf r})\cdot\bbox{\alpha}^{(s)}_k({\bf
 r})dv
=\int_{V_m}\bbox{\beta}_i^{(s)}({\bf r})\cdot\bbox{\alpha}^{(s)}_k({\bf
 r})dv.
\end{equation}
For simplicity, we introduce new conjugate variables $\tilde{q}_i$ and
 $\tilde{p}_i$ 
by canonical transformation $p_i=-\tilde{q}_i$
and $q_i=\tilde{p}_i$. If we define column vectors 
${\bf x}=^t(\{X^{(T)}_{ij}\}, \{X^{(s)}_{ij}\},
 \{\tilde{q}_i\},\{X^{(b)}_{ij}\})$
and  
${\bf p}=^t(\{P^{(T)}_{ij}\}, \{P^{(s)}_{ij}\},
 \{\tilde{p}_i\},\{P^{(b)}_{ij}\})$,
the total Hamiltonian can be written in a compact form as follows:
\begin{equation}
H=\frac{1}{2} \mbox{}^t{\bf p} T {\bf p} + \frac{1}{2}\mbox{}^t{\bf x} V {\bf
 x},
\label{eq:H4}
\end{equation}
where $T$ and $V$ are symmetric matrices. Since $T$ is positive definite, 
it has a decomposition $T=A^2$ by a symmetric regular matrix $A$. 
The positive semidefinite Hermite matrix 
$AVA$ can be diagonalized by an orthogonal matrix $U$ as 
${^tU} A VAU=D$, where $D$ is a diagonal matrix with no 
negative elements. A transformation to new bases 
${\bf X}={^t(}X_1,X_2,\ldots)$ and 
${\bf P}={^t(}P_1,P_2,\ldots)$ defined as 
${\bf X}= {^tU}A^{-1} {\bf x}$
 and 
${\bf p}=A^{-1}U{\bf P}$ 
is a canonical transformation since the commutation relations are
preserved:
\begin{eqnarray}
[X_i,P_j]&=&[\sum_k({^tU}A^{-1})_{ik}x_k,\sum_l({^tU}A)_{jl}p_k]
=\sum_{kl} ({^tU}A^{-1})_{ik}({^tU}A)_{jl}[x_k,p_l]
\nonumber \\
&=&i\hbar\sum_k({^tU}A^{-1})_{ik}(AU)_{kj}=i\hbar\delta_{ij}.
\end{eqnarray}
This transformation
diagonalizes the 
Hamiltonian (\ref{eq:H4}) into a sum of independent harmonic oscillators:
\begin{equation}
H=\frac{1}{2}{^t{\bf P}} {\bf P} + 
\frac{1}{2}{^t{\bf X}} D {\bf X}
=\sum_i \frac{1}{2}(P_i^2+\omega_i^2X_i^2)
+\sum_k \frac{1}{2}(P_k^{(b)2}+\omega_k^{(b)2}X_k^{(b)2}),
\label{eq:H5}
\end{equation}
where we have separated the contribution of bulk plasmons  
and suffixed their variables 
with superscript $(b)$ since they are decoupled from the others.

Now we second quantize Eq.\ (\ref{eq:H5}) to obtain
\begin{equation}
H=\sum_i\hbar\omega_i(a_i^\dagger a_i+\frac{1}{2})
+\sum_k\hbar\omega^{(b)}_k(a_k^{(b)\dagger} a_k^{(b)}+\frac{1}{2}),
\end{equation}
where we introduced creation and annihilation operators 
in the usual way. The energy of the ground state 
$\langle H\rangle_0$ is
\begin{equation}
\langle H\rangle_0=E_0+E^{(b)}_0\equiv \frac{1}{2}\sum_i\hbar\omega_i
+\frac{1}{2}\sum_k\hbar\omega_k^{(b)}.
\end{equation}
The Casimir force is derived from the change in the 
ground-state energy
with respect to
an infinitesimal displacement of the dielectric slabs. 
As we have seen,
the energy $E^{(b)}_0$ contributed by the bulk plasmons is independent of 
this displacement. Thus the Casimir force originates from $E_0$---the sum of 
zero-point energies of the harmonic-oscillator modes $X_i$. The relation
${\bf X}={^tU} A^{-1} {\bf x}$ implies that $X_i$ is a linear
 combination 
of $\tilde{q}_i$, $X^{(T)}_{ij}$, and $X^{(s)}_{ij}$. This means that 
$X_i$ represents a combined mode of the EM field and the collective
modes of charges in the dielectric. 
Therefore, in our approach, the Casimir force is attributable neither
to the change in zero-point energies of the genuine EM field nor to
that in zero-point energies of the genuine matter, but to that in 
{\it zero-point energies of the combined matter-field modes}.

\section{EQUIVALENCE TO MAXWELL EQUATIONS}
\label{sec:max}
In this section, we show that the diagonalization of the Hamiltonian
(\ref{eq:H3}) amounts to solving the Maxwell equations associated with 
a proper dielectric response function 
$\epsilon(\omega)$ under  
appropriate boundary conditions.

Consider the equations of motion derived from the Hamiltonian (\ref{eq:H3}).
After eliminating $q_i$, $P^{(T)}_{ij}$, $P^{(s)}_{ij}$, and $P^{(b)}_{ij}$,
these equations read
\begin{eqnarray}
\ddot{p}_i  &=& -\Omega_i^2 p_
+\sum_{kj}\ddot{X}^{(T)}_{kj}\int_{V_m}
\bbox{\alpha}_i^{(T)}\cdot\bbox{\beta}_k^{(T)}dv
+\sum_{kj}\ddot{X}^{(s)}_{kj}\int_{V_m}
\bbox{\alpha}_i^{(T)}\cdot\bbox{\beta}_k^{(s)}dv,
\label{eq:p}
\\
\ddot{X}^{(T)}_{kj}&=&-\omega_{0j}^2 X^{(T)}_{kj}
-\omega_{pj}^2\sum_ip_i\int_{V_m} 
\bbox{\beta}_k^{(T)}\cdot\bbox{\alpha}_i^{(T)}dv,
\label{eq:T}
\\
\ddot{X}^{(s)}_{kj}&=&-\omega_{0j}^2 X^{(s)}_{kj}
-\omega_{pj}^2\sum_ip_i\int_{V_m}
\bbox{\beta}_k^{(s)}\cdot\bbox{\alpha}_i^{(T)}dv
-\omega_{pj}^2\sum_l K_{kl} \sum_{j^\prime}
X^{(s)}_{lj^\prime},
\\
\ddot{X}^{(b)}_{kj}&=&-\omega_{0j}^2 X^{(b)}_{kj}
-\omega_{pj}^2 \sum_{j^\prime}X^{(b)}_{kj^\prime}.
\label{eq:b}
\end{eqnarray}
The diagonalization of the Hamiltonian 
is equivalent to finding the solution to the above 
set of equations that oscillates at frequency $\omega$. For such 
a solution, the second-order time derivative may be replaced by $-\omega^2$. 
Equation (\ref{eq:p}) then reduces to
\begin{equation}
p_i = \frac{\omega^2}{\omega^2-\Omega_i^2}
\left(
\sum_{k}X^{(T)}_{k}\int_{V_m}
\bbox{\alpha}_i^{(T)}\cdot\bbox{\beta}_k^{(T)}dv
+\sum_{k}X^{(s)}_{k}\int_{V_m}
\bbox{\alpha}_i^{(T)}\cdot\bbox{\beta}_k^{(s)}dv 
\right),
\label{eq:p2}
\end{equation}
where we used the notation
\begin{equation}
X_k^{(M)}\equiv\sum_jX_{kj}^{(M)} \;\;\;\; (M=T,s,b).
\end{equation}
From the remaining equations (\ref{eq:T})--(\ref{eq:b}), 
we obtain after summation over the suffix $j$
\begin{eqnarray}
X^{(T)}_k &=& (1-\epsilon(\omega))\sum_ip_i\int_{V_m}
\bbox{\beta}_k^{(T)}\cdot\bbox{\alpha}_i^{(T)}dv,
\label{eq:T2}
\\
X^{(s)}_k &=& (1-\epsilon(\omega))
\left[
\sum_ip_i\int_{V_m}
\bbox{\beta}_k^{(s)}\cdot\bbox{\alpha}_i^{(T)}dv
+\sum_l X^{(s)}_l\int_{V_m}
\bbox{\beta}_k^{(s)}\cdot\bbox{\alpha}_l^{(s)}dv
\right],
\\
X^{(b)}_k &=& (1-\epsilon(\omega))X^{(b)}_k,
\label{eq:b2}
\end{eqnarray}
where we have introduced the dielectric response 
function $\epsilon(\omega)$ defined as 
\begin{equation}
\epsilon(\omega)\equiv 1-\sum_j\frac{\omega_{pj}^2}{\omega^2-\omega_{0j}^2}.
\end{equation}

Let us introduce the electric field ${\bf E}({\bf r})$ defined as
\begin{equation}
{\bf E}({\bf r})\equiv
-\frac{1}{\sqrt{\epsilon_0}}
\left(
\sum_i p_i\bbox{\alpha}^{(T)}_i({\bf r})
+\sum_k X^{(s)}_k\bbox{\alpha}^{(s)}_k({\bf r})
+\sum_l X^{(b)}_l\bbox{\alpha}^{(b)}_l({\bf r})
\right).
\label{eq:defE}
\end{equation}
Then it is easy to show that Eqs. (\ref{eq:T2})--(\ref{eq:b2}) have the same
 form 
as follows:
\begin{equation}
X^{(M)}_k=\sqrt{\epsilon_0}(\epsilon(\omega)-1)
\int_{V_m}\bbox{\beta}_k^{(M)}\cdot{\bf E}dv
\;\;\;\;\;\; (M=T,s,b).
\label{eq:Tsb}
\end{equation}
Substituting this into Eq. (\ref{eq:p2}) and noting the completeness
 relation
\begin{equation}
\sum_i\bbox{\beta}_i^{(T)}({\bf r})\bbox{\beta}_i^{(T)}({\bf r}^\prime)
+\sum_k\bbox{\beta}_k^{(s)}({\bf r})\bbox{\beta}_k^{(s)}({\bf r}^\prime)
+\sum_l\bbox{\beta}_l^{(b)}({\bf r})\bbox{\beta}_l^{(b)}({\bf r}^\prime)
=\tensor{1}\delta^3({\bf r}-{\bf r}^\prime),
\label{eq:comp}
\end{equation}
where $\tensor{1}$ denotes the identity matrix, we obtain
\begin{equation}
p_i=\frac{\sqrt{\epsilon_0}(\epsilon(\omega)-1)\omega^2}{\omega^2-\Omega_i^2}
\int_{V_m}\bbox{\alpha}_i^{(T)}\cdot{\bf E}dv.
\label{eq:47}
\end{equation}
Substituting Eqs.\ (\ref{eq:Tsb})
and (\ref{eq:47}) into Eq.\ (\ref{eq:defE}) yields the
 self-consistent condition
for the electric field ${\bf E}({\bf r)}$:
\begin{eqnarray}
{\bf E}({\bf r)}&=&[1-\epsilon(\omega)]
\left(
\sum_i\frac{\omega^2}{\omega^2-\Omega_i^2}
\bbox{\alpha}^{(T)}_i({\bf r})
\int_{V_m}\bbox{\alpha}^{(T)}_i({\bf r}^\prime)\cdot{\bf E}({\bf
 r}^\prime)dv^\prime
\right.
\nonumber \\
& & \left.
+\sum_k\bbox{\alpha}^{(s)}_k({\bf r})
\int_{V_m}\bbox{\beta}^{(s)}_k({\bf r}^\prime)\cdot{\bf E}({\bf
 r}^\prime)dv^\prime
+\sum_l\bbox{\alpha}^{(b)}_l({\bf r})
\int_{V_m}\bbox{\beta}^{(b)}_l({\bf r}^\prime)\cdot{\bf E}({\bf
 r}^\prime)dv^\prime
\right).
\label{eq:eonly}
\end{eqnarray}
This equation can be simplified if we convert it into a pair of equations as
 follows.
First, by taking the divergence of both sides we obtain
\begin{eqnarray}
\bbox{\nabla}\cdot{\bf E}({\bf r)}&=&[1-\epsilon(\omega)]\left(
\sum_k\bbox{\nabla}\cdot\bbox{\alpha}^{(s)}_k({\bf r})
\int_{V_m}\bbox{\beta}^{(s)}_k({\bf r}^\prime)\cdot{\bf E}({\bf
 r}^\prime)dv^\prime
\right. \nonumber \\
&&\left. +\sum_l\bbox{\nabla}\cdot\bbox{\alpha}^{(b)}_l({\bf r})
\int_{V_m}
\bbox{\beta}^{(b)}_l({\bf r}^\prime)\cdot{\bf E}({\bf r}^\prime)dv^\prime
\right).
\label{eq:49}
\end{eqnarray}
Substituting Eq.\ (\ref{eq:defalpha}) into Eq.\ (\ref{eq:49}),
we obtain
\begin{eqnarray}
\bbox{\nabla}\cdot{\bf E}({\bf r)}&=&[1-\epsilon(\omega)]\bbox{\nabla}\cdot\left(
\sum_k\bbox{\beta}^{(s)}_k({\bf r})
\int_{V_m}\bbox{\beta}^{(s)}_k({\bf r}^\prime)\cdot{\bf E}({\bf
 r}^\prime)dv^\prime
\right. \nonumber \\
&&\left. +\sum_l\bbox{\beta}^{(b)}_l({\bf r})
\int_{V_m}
\bbox{\beta}^{(b)}_l({\bf r}^\prime)\cdot{\bf E}({\bf r}^\prime)dv^\prime
\right) \Theta({\bf r}),
\label{eq:dive}
\end{eqnarray}
where $\Theta({\bf r})$ is the unit step function defined by
\begin{equation}
\Theta({\bf r})=\left\{
 \begin{array}{ll}
1 \;\;\;\;\; & \mbox{in $V_m$}, \\
0 & \mbox{otherwise}.
\end{array}
\right. 
\end{equation}
Using the completeness relation (\ref{eq:comp}) and noting that 
$\bbox{\beta}^{(T)}_i({\bf r})\Theta({\bf r})$ is a transverse function
in $V$, Eq. (\ref{eq:dive}) is simplified, giving
\begin{equation}
\bbox{\nabla}\cdot{\bf E}({\bf r)}=[1-\epsilon(\omega)]\bbox{\nabla}\cdot
[{\bf E}({\bf r)}\Theta({\bf r})].
\label{eq:div}
\end{equation}

On the other hand, by operating $(\omega^2-c^2\mbox{rotrot})$ on both sides 
of Eq. (\ref{eq:eonly}) and using Eq.\ (\ref{eq:Omega}), we have
\begin{eqnarray}
(\omega^2-c^2\mbox{rotrot}){\bf E}({\bf r)}&=&[1-\epsilon(\omega)]\omega^2
\left(
\sum_i
\bbox{\alpha}^{(T)}_i({\bf r})
\int_{V_m}\bbox{\alpha}^{(T)}_i({\bf r}^\prime)\cdot{\bf E}({\bf
 r}^\prime)dv^\prime
\right.
\nonumber \\
& & \left.
+\sum_k\bbox{\alpha}^{(s)}_k({\bf r})
\int_{V_m}\bbox{\beta}^{(s)}_k({\bf r}^\prime)\cdot{\bf E}({\bf
 r}^\prime)dv^\prime
+\sum_l\bbox{\alpha}^{(b)}_l({\bf r})
\int_{V_m}\bbox{\beta}^{(b)}_l({\bf r}^\prime)\cdot{\bf E}({\bf
 r}^\prime)dv^\prime
\right).
\end{eqnarray}
Here, using (\ref{eq:alphal}) and (\ref{eq:comp}), we obtain
\begin{eqnarray}
& & \sum_i
\bbox{\alpha}^{(T)}_i({\bf r})
\bbox{\alpha}^{(T)}_i({\bf r}^\prime)
+\sum_k\bbox{\alpha}^{(s)}_k({\bf r})
\bbox{\beta}^{(s)}_k({\bf r}^\prime)
+\sum_l\bbox{\alpha}^{(b)}_l({\bf r})
\bbox{\beta}^{(b)}_l({\bf r}^\prime)
\nonumber \\
&=&
\sum_k\Theta({\bf r})\bbox{\beta}^{(s)}_k({\bf r})
\bbox{\beta}^{(s)}_k({\bf r}^\prime)
+\sum_l\Theta({\bf r})\bbox{\beta}^{(b)}_l({\bf r})
\bbox{\beta}^{(b)}_l({\bf r}^\prime)
\nonumber \\
&&+\sum_i
\bbox{\alpha}^{(T)}_i({\bf r})
\bbox{\alpha}^{(T)}_i({\bf r}^\prime)
-\sum_{ik}\bbox{\alpha}^{(T)}_i({\bf r})
\int_{V_m}\bbox{\alpha}^{(T)}_i({\bf r}^{\prime\prime})\cdot
\bbox{\beta}^{(s)}_k({\bf r}^{\prime\prime})dv^{\prime\prime}
\bbox{\beta}^{(s)}_k({\bf r}^\prime)
\nonumber \\
&=&\sum_k\Theta({\bf r})\bbox{\beta}^{(s)}_k({\bf r})
\bbox{\beta}^{(s)}_k({\bf r}^\prime)
+\sum_l\Theta({\bf r})\bbox{\beta}^{(b)}_l({\bf r})
\bbox{\beta}^{(b)}_l({\bf r}^\prime)
\nonumber \\
&&+\sum_{ik}\bbox{\alpha}^{(T)}_i({\bf r})
\int_{V_m}\bbox{\alpha}^{(T)}_i({\bf r}^{\prime\prime})\cdot
\bbox{\beta}^{(T)}_k({\bf r}^{\prime\prime})dv^{\prime\prime}
\bbox{\beta}^{(T)}_k({\bf r}^\prime)
\nonumber \\
&=&\Theta({\bf r})\tensor{1}\delta^3({\bf r}-{\bf r}^\prime),
\end{eqnarray}
where we have used the fact that $\Theta({\bf r}^{\prime\prime})
\bbox{\beta}^{(T)}_k({\bf r}^{\prime\prime})$ can be expanded with 
$\bbox{\alpha}^{(T)}_i({\bf r}^{\prime\prime})$. We thus obtain
\begin{equation}
(\omega^2-c^2\mbox{rotrot}){\bf E}({\bf r})=[1-\epsilon(\omega)]\omega^2
{\bf E}({\bf r})\Theta({\bf r}).
\label{eq:rot}
\end{equation}

If we define the electric displacement ${\bf D}({\bf r})$ as 
${\bf D}({\bf r})\equiv\epsilon_0\{1+[\epsilon(\omega)
-1]\Theta({\bf r})\}{\bf E}({\bf r})$, we find that 
Eqs.\ (\ref{eq:div}) and (\ref{eq:rot}) are identical to the Maxwell 
equations, namely,
\begin{equation}
\bbox{\nabla}\cdot{\bf D}({\bf r})=0
\end{equation}
and 
\begin{equation}
\bbox{\nabla}\times(\bbox{\nabla}\times{\bf E}({\bf r}))=\frac{\omega^2}{\epsilon_0c^2}
{\bf D}({\bf r}).
\end{equation}
Our theory developed in Sec.\ \ref{sec:formal} therefore
 yields the same sequence
of eigenvalues $\omega_i$ as Schram's method\cite{schram73} and 
eigenvalues of the bulk 
plasmon modes that are not relevant to the Casimir force.
Our theory thus
gives the identical result for the
 Casimir
force: for example, when the interface $S_m$ consists of 
a pair of 
parallel planes with a distance $d$, the force per area is
calculated to be
\begin{equation}
F=-\frac{\hbar}{2\pi^2c^3}\int^{\infty}_0d\xi\int^{\infty}_1dp
p^2\xi^3\left[
\left\{
\left(
\frac{s+p\epsilon(i\xi)}{s-p\epsilon(i\xi)}
\right)^2e^{2p\xi d/c}-1
\right\}^{-1}
+
\left\{
\left(
\frac{s+p}{s-p}
\right)^2e^{2p\xi d/c}-1
\right\}^{-1}
\right],
\end{equation}
where $s\equiv \sqrt{p^2+\epsilon{i\xi}-1}$.

\section{Field theories and matter theories of the Casimir force}
\label{sec:previous}

In the preceding sections, we have developed a theory of 
the Casimir force by starting from the Lagrangian that describes both the 
EM field and the collective modes of charges in the matter. 
In this theory, the Casimir effect is attributed to a change in 
zero-point energies of the combined modes of the EM field and 
the matter. Thus, in our theory, both field and matter 
contribute to the Casimir force. 
The system considered here, however, has
one asymmetry between the field and matter, which is obvious from 
Fig. \ref{fig:1}. The asymmetry originates from the property of the 
interface $S_m$ in Fig.\ \ref{fig:1}. It confines 
the matter in one side, but imposes no boundary condition 
on the EM field.  
It is difficult to imagine an opposite situation.  
That is, we can 
exclude the matter from some region, but we
cannot clear away the EM vacuum.
As will be discussed below, it is this asymmetry 
that makes some of the existing theories favor the interpretation 
of the Casimir force as arising from 
zero-point EM 
energies (field theories), 
and others favor the interpretation as due to zero-point fluctuations 
of the matter 
(matter theories).

First, we consider a field theory of the Casimir force,
by which we refer to the 
scheme of solving the Maxwell equations under appropriate boundary 
conditions. The relation between our theory
and the field theory has already been discussed in Sec.\ \ref{sec:max}. 
There we have shown 
that both theories actually calculate the zero-point energies of 
the same set of combined matter-field modes. 
Nevertheless these energies 
are sometimes incorrectly  
identified  as  zero-point energies 
of the genuine EM field, presumably because the
normal modes can be specified by only invoking the Maxwell equations
without 
any reference to the state of the matter. The degrees of freedom of 
the matter are embedded in the dielectric response function from the 
very beginning 
in the field theory of the Casimir force
and are therefore not manifest. In addition,
in the simplest case of the two perfectly conducting plates, 
it would be easy to 
forget the matter because it occupies only the thin surface 
and 
the whole mode volume is filled with the pure EM field. 
As a matter of fact, however, 
within this thin surface region there exist rather complicated surface
modes, as described in Sec.\ \ref{sec:formal}, which conspire to shield
the EM field, producing the Casimir effect.
In contrast, these normal modes can hardly be 
looked upon as genuine matter oscillations because 
it is difficult to construct a simple wave equation that contains 
only the matter variables when the system includes regions where no matter
exists. 

A matter theory of the Casimir force, which is based on zero-point 
fluctuations of the matter, is 
 formulated by Milonni and Shih\cite{milonni92} in terms of 
conventional quantum electrodynamics.
Their scheme looks somewhat complicated, but the essential ingredient of
the theory is the use of
second-order perturbation theory in 
calculating the self-energy of linearly interacting harmonic 
oscillators.
Here we will present the bare essentials of this theory 
by taking up the simplest example 
of a pair of
interacting harmonic oscillators.

The Hamiltonian for the system is written as
\begin{equation}
H=H_{0}+V,
\end{equation}
where 
\begin{eqnarray}
H_{0}&=&\hbar\omega_a\left(\hat{a}^\dagger\hat{a}+\frac{1}{2}\right)
+\hbar\omega_b\left(\hat{b}^\dagger\hat{b}+\frac{1}{2}\right),
\\
V&=&v_+\hat{a}^\dagger\hat{b}^\dagger
+v_-\hat{a}^\dagger\hat{b}
+v^*_+\hat{b}\hat{a}
+v^*_-\hat{b}^\dagger\hat{a},
\label{eq:v}
\end{eqnarray}
and the interaction $V$ is assumed to be small.
In this case, the first nonvanishing correction to the 
ground-state energy is of second order in $V$ and is written as 
$\Delta E^{(2)}=\langle g^{(0)}|V|g\rangle$, where the state
 $|g^{(0)}\rangle$ is 
the unperturbed ground state and $|g\rangle$ is the ground state, 
which is correct up
 to 
first order in $V$. 
Instead of using a time-independent perturbation theory, we
 can also 
obtain $|g\rangle$ by a time-dependent perturbation theory with the
 Hamiltonian 
$H_{0}+Ve^{\gamma t}$, where the limit $\gamma\rightarrow +0$ is implied. 
Hence
\begin{eqnarray}
|g\rangle&=&\left(1-\frac{i}{\hbar}\int_{-\infty}^0dt^\prime
e^{iH_0t^\prime}Ve^{\gamma t^\prime}e^{-iH_0t^\prime}\right)|g^{(0)}\rangle
\nonumber \\
&\equiv & U |g^{(0)}\rangle.
\end{eqnarray}
Up to second order, we have
\begin{eqnarray}
\Delta E^{(2)}&=&\langle g^{(0)} |VU|g^{(0)}\rangle
=\langle g^{(0)} |U^\dagger V|g^{(0)}\rangle
\nonumber \\
&=&\frac{1}{2}\langle g^{(0)} |U^\dagger VU|g^{(0)}\rangle.
\label{eq:deltae}
\end{eqnarray}
This corresponds to the interaction energy assumed in Ref.~\cite{milonni92}.
The following relations are useful for the following calculation:
\begin{eqnarray}
U^\dagger \hat{a} U &=& \hat{a}-\frac{v_+}{\omega_a+\omega_b}\hat{b}^\dagger
-\frac{v_-}{\omega_a-\omega_b}\hat{b},
\label{eq:uau}
\\
U^\dagger \hat{b} U &=& \hat{b}-\frac{v_+}{\omega_a+\omega_b}\hat{a}^\dagger
-\frac{v_-}{\omega_b-\omega_a}\hat{a}.
\label{eq:ubu}
\end{eqnarray}

Since the interaction $V$ consists of the products of commuting operators, 
changing the ordering of 
operators in $V$ should not affect the value of 
 $\Delta E^{(2)}$. Let us take one particular
ordering of
\begin{equation}
V=v_+\hat{a}^\dagger\hat{b}^\dagger
+v_-\hat{a}^\dagger\hat{b}
+v^*_+\hat{b}\hat{a}
+v^*_-\hat{b}^\dagger\hat{a},
\end{equation}
and substitute Eqs.\ (\ref{eq:uau}) and (\ref{eq:ubu}) 
directly into $U^\dagger VU$ without changing the order of operators. 
In the resulting 
expression, the operators 
for mode $a$ appear in normal order, and therefore do not 
contribute to the expectation value in the vacuum. 
Thus, $\Delta E^{(2)}$ in Eq.\ (\ref{eq:deltae})
is expressed only in terms of zero-point
 fluctuations of mode $b$ as
\begin{equation}
\Delta E^{(2)}=-\frac{|v_+|^2}{\omega_a+\omega_b}
\langle g^{(0)} |\hat{b}\hat{b}^\dagger|g^{(0)}\rangle.
\end{equation}

If we take another choice of operator ordering for $V$, 
\begin{equation}
V=v_+\hat{b}^\dagger\hat{a}^\dagger
+v_-\hat{a}^\dagger\hat{b}
+v^*_+\hat{a}\hat{b}
+v^*_-\hat{b}^\dagger\hat{a},
\end{equation}
the same procedure leads to
\begin{equation}
\Delta E^{(2)}=-\frac{|v_+|^2}{\omega_a+\omega_b}
\langle g^{(0)} |\hat{a}\hat{a}^\dagger|g^{(0)}\rangle.
\end{equation}
In this case, the correction is attributable to  
quantum fluctuations of mode
 $a$.
The second-order correction to 
the ground-state energy of
a pair of linearly coupled harmonic oscillators can therefore be
 attributed
solely to zero-point fluctuations of either one of the oscillators.

Now, back to the original problem of the Casimir force. 
We notice\cite{milonni92} 
that the force is derived from the change in the energy 
of the entire system when we add
 atoms in an infinitesimally thin
layer next to the boundary $S_m$\cite{foot1}.
 The dominant contribution to 
this change arises from 
the interaction energy between the added atoms and the field
 modes, either
of which may be described as a set of harmonic oscillators.
Thus, as in the simple system discussed above, we can attribute the change
 in energy
to the fluctuations of the added atoms alone, provided that
an appropriate operator ordering is chosen (matter theory).
We emphasize that the logical consistency of this interpretation
hinges heavily on  second-order 
perturbation theory as applied to a system of linearly coupled
harmonic oscillators (linear-response theory). If 
effects of higher-order interactions are not 
negligible, it would be nontrivial to construct either genuine
matter theory or field theory that is consistent with each other.

If we choose another appropriate operator ordering, the Casimir force 
may be interpreted as arising from
quantum fluctuations of the EM field
 modes, and we obtain a kind of field theory of the Casimir force. 
But the machinery is similar to 
the field theories described earlier, that is, 
the field modes are actually
not the genuine EM field modes but 
the combined 
modes of the EM field and the matter.

Second-order perturbation theory (or linear-response theory) thus
allows the interpretation
of the Casimir force as arising from quantum 
fluctuations of the genuine 
matter alone but not from those of
 the genuine
EM field alone. The reason for this asymmetry 
lies again in the property of the boundary. That is,
an infinitesimal
displacement of the boundary only affects 
the mode functions of the genuine matter but has no
 direct effect on the mode functions of 
the genuine EM field.

\section{Summary}
\label{sec:conclusion}

The origin of the Casimir force has usually been attributed to zero-point
 fluctuations of the
EM field, and sometimes to those of the matter. 
We have formulated the problem by starting  
from the Lagrangian that describes the EM field, collective
modes of the matter, and their interaction.
This approach makes it clear that the Casimir effect,
in actual fact, arises from the change in
zero-point energies of certain combined modes
of the transverse EM field, the transverse motion of charges, 
and the surface plasmons. The zero-point energy of the 
entire system is therefore
 contributed by both
the EM field and the matter.

The Casimir force arises not from the zero-point energy {\it per se}, 
but from its change with respect to
a virtual infinitesimal 
displacement of the plates. This change alters the mode functions
of the matter alone, and does not affect those of the EM field.
Because of this asymmetry, it is possible to construct a genuine
matter theory 
but it is difficult to conceive a genuine field theory.

A most natural derivation of the Casimir force seems to be 
to calculate the zero-point energy first and then take its derivative.
The original derivation by Casimir follows this approach. In this
 type of 
approach, both field and matter 
contribute to the zero-point energy as in our
 theory.
Nevertheless, one encounters descriptions in the literature
to the effect that the Casimir force is the hallmark of 
zero-point energies of the genuine EM field.
Perhaps the main reason for this (strictly speaking incorrect)
recognition lies in 
the fact that while the system has the region where only the EM field
 (including the vacuum) 
exists, there is no region where only the matter exists. 
This also reflects the above-mentioned  asymmetry between the 
field and the matter.

\section*{Acknowledgment}

M.U. acknowledges support by the Core Research for Evolutional Science and
Technology (CREST) of the Japan Science and Technology Corporation.


\begin{thebibliography}{1}

\bibitem{casimir48}
H.~B.~G. Casimir, Proc.~K.~Ned.~Akad.~Wet. {\bf 51},  793  (1948).

\bibitem{lifshitz56}
E.~M. Lifshitz, Zh.~Eksp.~Teor.~Fiz. {\bf 29}, 94 (1955)
[Sov.~Phys.~JETP {\bf 2},  73  (1956)].

\bibitem{kampen68}
N.~G. van Kampen, B.~R.~A. Nijboer, and K. Schram, Phys.~Lett. {\bf 26A}, 
 307
  (1968).

\bibitem{schram73}
K. Schram, Phys.~Lett. {\bf 43A},  282  (1973).

\bibitem{boyer68}
T.~H. Boyer, Phys.~Rev. {\bf 174},  1764  (1968).

\bibitem{sparnaay58}
M.~J. Sparnaay, Physica (Utrecht) {\bf 24}, 751 (1958).

\bibitem{tabor68}
D. Tabor and R.~H.~S. Winterton, Nature (London) {\bf 219}, 1120 (1968).

\bibitem{israelachvili72}
J.~N. Israelachvili and D. Tabor, Proc.~R.~Soc.~London A {\bf 331}, 19 (1972).

\bibitem{lamoreaux97}
S.~K.~Lamoreaux, Phys.~Rev.~Lett. {\bf 78}, 5 (1997).

\bibitem{polder48}
H.~B.~G. Casimir and D. Polder, Phys.~Rev. {\bf 73}, 360 (1948).


\bibitem{sukenik93}
C.~I. Sukenik, M.~G. Boshier, D. Cho, V. Sandoghdar, and E.~A. Hinds,
Phys.~Rev.~Lett. {\bf 70}, 560 (1993).

\bibitem{mehra67}
J. Mehra, Physica (Utrecht) {\bf 37}, 145 (1967).

\bibitem{brown69}
L.~S. Brown and G.~J. Maclay, Phys.~Rev. {\bf 184}, 1272 (1969).

\bibitem{schwinger78}
J. Schwinger, L.~L. DeRaad, and K.~A. Milton, Ann. Phys. (N.Y.) {\bf 115}, 
 1 (1978).

\bibitem{xinwei97}
X. Kong and F. Ravndal,
Phys.~Rev.~Lett. {\bf 79}, 545 (1997).

\bibitem{plunien85}
G. Plunien, B. M\"uller, and W. Greiner,
Phys.~Rep. {\bf 134}, 87 (1986).

\bibitem{mostepanenko88}
V.~M. Mostepanenko and N.~N. Trunov, Usp.~Fiz.~Nauk. {\bf 156}, 385 (1988)
[Sov.~Phys.~Usp. {\bf 31}, 965 (1988)].

\bibitem{elizalde91}
E. Elizalde and A. Romeo, Am.~J.~Phys. {\bf 59}, 711 (1991).

\bibitem{spruch96}
L. Spruch, Science {\bf 272}, 1452 (1996).

\bibitem{schwinger75}
J. Schwinger, Lett. Math. Phys. {\bf 1},  43  (1975).

\bibitem{milonni92}
P.~W. Milonni and M.~-L. Shih, Phys. Rev. A {\bf 45},  4241  (1992).

\bibitem{foot1}
Strictly speaking, when we add some amounts of atoms on one surface of
the dielectric slab, we should remove the same amount of atoms from the
other surface so that the volume of the slab is kept constant; otherwise
the bulk plasmons would also contribute to the change in energy. We should 
take into account both effects 
of adding and removing atoms in evaluating the change in 
energy.




\end{thebibliography}
\end{document}